\documentclass[12pt]{article}
\usepackage{amsmath,amsthm,amscd,amsfonts,amssymb}
\usepackage{latexsym}
\usepackage{mathabx}
\usepackage[usenames]{color}
\usepackage{comment}

\newcommand{\NN}{\mathbb N}
\newcommand{\PP}{\mathbb P}
\newcommand{\RR}{\mathbb R}

\newcommand{\ZZ}{\mathbb Z}
\newcommand{\EE}{\mathbb E}
\newcommand{\LL}{\mathbb L}

\renewcommand{\aa}{\mathcal A}
\newcommand{\bb}{\mathcal B}
\newcommand{\cc}{\mathcal C}

\newcommand{\ff}{\mathcal F}

\newcommand{\zz}{\mathcal Z}
\newcommand{\WW}{\mathcal S}
\newcommand{\xx}{\mathcal X}

\newcommand{\Ee}{\mathcal E}
\newcommand{\Ll}{\mathcal L}
\newcommand{\da}{d_{1}}
\newcommand{\db}{d_{2}}
\newcommand{\supp}{\rm{supp}\,}
\newtheorem{thm}{Theorem}[section]
\newtheorem{lemma}[thm]{Lemma}
\newtheorem{cor}[thm]{Corollary}
\newtheorem{prop}[thm]{Proposition}
\newtheorem{defin}[thm]{Definition}
\newtheorem{rem}[thm]{Remark}
\newtheorem{hyp}[thm]{Hypothesis}
\newtheorem{exam}[thm]{Examples}

\def\half{\frac{1}{2}}

\newcommand{\dist}{\rm{dist}}
\newcommand{\tr}{\rm{tr\,}}

\newenvironment{proofn}[1][]{\noindent\textbf{Proof}#1\textbf{:} }{\ \hfill \rule{0.5em}{0.5em}\\[2mm]}

\begin{document}
\author{ Werner Kirsch\\ Institut f\"ur Mathematik und Informatik\\ FernUniversit\"at in Hagen \\
58097 Hagen, Germany \\ email: werner.kirsch@fernuni-hagen.de \\
M. Krishna  \\ Ashoka University, Plot 2, Rajiv Gandhi Education City\\
Rai, Haryana 131029 India\\ email: krishna.maddaly@ashoka.edu.in  \\
}
\title{Quantum Lattice Wave Guides with Randomness - Localisation and Delocalisation}
\maketitle
\begin{abstract}
In this paper We consider Schr\"{o}dinger operators  on $M\times\ZZ^{\db}$, with $M=\{M_{1},\ldots,M_{2}\}^{\da}$  (`quantum wave guides') with a `$\Gamma$-trimmed' random potential, namely a potential which vanishes outside a subset $\Gamma$ which is periodic with respect to a sub lattice.

We prove that (under appropriate assumptions) for strong disorder these
operators have \emph{pure point spectrum } outside the set $\Sigma_{0}=\sigma(H_{0,\Gamma^{c}})$ where
$H_{0,\Gamma^{c}} $ is the free (discrete) Laplacian on the complement $\Gamma^{c} $ of $\Gamma $.
We also prove that the operators have some \emph{absolutely continuous spectrum} in an energy region $\Ee\subset\Sigma_{0}$.
Consequently, there is a mobility edge for such models.

We also consider the case $-M_{1}=M_{2}=\infty$, i.~e.~ $\Gamma $-trimmed operators on $\ZZ^{d}=\ZZ^{\da}\times\ZZ^{\db}$. Again, we prove localisation outside $\Sigma_{0} $ by showing
exponential decay of the Green function $G_{E+i\eta}(x,y) $ uniformly in $\eta>0 $.

For \emph{all} energies $E\in\Ee$
we prove that the Green's function $G_{E+i\eta} $ is \emph{not} (uniformly) in $\ell^{1}$ as $\eta$ approaches $0$. This implies that neither the fractional moment method nor
multi scale analysis \emph{can} be applied here.
\end{abstract}

\section{Introduction}
Quantum Waveguides are quantum mechanical structures which are confined in certain spaces dimensions,
but unconfined in others. The last decades showed a growing interest on these systems in the mathematical literature. The recent book \cite{ExnerK} by Exner and Kova\v{r}\'{\i}k  gives an overview on the state of art (as of 2015)
as well as an extensive list of references on waveguides.

A quantum waveguide with the simplest geometry is given by a particle in a ($k$-dimensional) strip in $\ZZ^{k+m}$ or $\RR^{k+m}$. Other examples are tubes or wires which are bended  or twisted (see for example the discussion in Krej\v{c}i\v{r}\'{\i}k \cite{krej}). Of particular interest are waveguides with randomness
either in the geometry of the system or in the potential energy.

In this paper we consider waveguides with a simple geometry, namely on a strip in $\ZZ^{d}$, for example on
\begin{align}\label{eq:xx}
\xx~=~\{M_1p,M_1+1\ldots, M_2 p-1 \}^{d_{1}}\times\ZZ^{\db} ,\qquad \da+\db=d, p\geq 2\,,
\end{align}
with a random potential $V_{\omega} $. The potential we consider is `sporadic' or `$\Gamma $-trimmed', in the sense that $V_{\omega}(x)=0$ for
lattice points $x\not\in\Gamma$. Here, $\Gamma $ is an $\LL$- periodic subset of the strip for
a sub lattice $\LL$.
In example \eqref{eq:xx}, we may choose for instance
\begin{align}\label{eq:Gamma}
\Gamma=\{M_1 p, (M_1+1)p,\ldots,M_2 p\}^{d_1}\times \ZZ^{d_2}
\end{align}
For $x\in\Gamma$ the potentials are independent and identically distributed. Random operators $H_{\omega}$ with such potentials are called `$\Gamma $-trimmed'.

Spectral theory for trimmed Anderson models (i.~e. on $\ZZ^{d} $) was done in the PhD-thesis of Obermeit \cite{JO} and the papers of  Rojas-Molina \cite{CR}, Elgart-Klein \cite{EK}, Elgart-Sodin \cite{ES} and Kirsch-Krishna \cite{KK3}. The present paper was inspired by \cite{ES}.

We show that models as in \eqref{eq:xx}, \eqref{eq:Gamma} have a mobility edge (or rather mobility edges). The measure theoretical nature
of spectrum depends on the energy region. More precisely, denote by $H_{0,\Gamma^{c}}$ the free
(discrete) Laplacian on the set $ \Gamma^{c}$. Outside of the spectrum $\Sigma_{0}:=\sigma(H_{0,\Gamma^{c}}) $
the operator $H_{\omega}$ has \emph{dense point spectrum} for high enough disorder. On the other hand, we prove that $H_{\omega}$ has some \emph{absolutely continuous spectrum} inside $\sigma(H_{0,\Gamma^{c}})$ regardless of the strength (or even existence) of the randomness.

The absolutely continuous spectrum comes from the existence of canonical extended states. More precisely,
in an energy regions inside $\Sigma_0$ we find periodic solutions
of the free Schr\"odinger equation which vanish on the set $\Gamma$. Hence these functions
solve the Schr\"odinger equation with the random (or arbitrary) potential on $\Gamma$ as well.

To prove pure point spectrum we employ the multiscale analysis (see e.~g. Dreifus-Klein \cite{vDK} or Disertori et.al. \cite{Diss3K} and references given there). The critical ingredient in our case is a \emph{Wegner Estimate}. To prove this estimate we use in an essential
way that we work outside the spectrum $\sigma(H_{0,\Gamma^{c}})$. In fact, the estimate blows up when we approach
$\sigma(H_{0,\Gamma^{c}})$.

From the multiscale bounds not only pure point spectrum follows but also dynamical localisation (see Damanik-Stollmann
\cite{DamanikS}).

We also consider the case $-M_{1}=M_{2}=\infty$, in other words a $\Gamma $-trimmed potential on $\ZZ^{d}=\ZZ^{\da}\times\ZZ^{\db}$. Again we use multiscale analysis outside $\Sigma_{0}$ for high enough disorder, which implies uniform exponential decay of the Green function and existence of pure
point spectrum.

Inside $\Sigma_{0}$ we prove that the Green function
$G_{E+i \varepsilon}(x,y)$ is not only not uniformly exponentially decaying, but that for $x\notin\Gamma$ even
\begin{align}
   \sup_{\varepsilon\nearrow 0}\,\sum_{y\in\ZZ^{d}}\,|G_{E+i\varepsilon}(x,y)|~=~\infty\,.
\end{align}
Consequently, we have a `phase transition' at  $\Sigma_{0}=\sigma(H_{0,\Gamma^{c}})$ which
manifests itself in the behaviour of the Green function.

For the Anderson model (with full randomness) it is expected that in higher dimension
there is a mobility edge, namely a
transition from pure point spectrum to absolutely continuous spectrum depending on the energy range
and the strength of the disorder. All that is rigorously known (on $\ZZ^{d}$) is the existence
of pure point spectrum (see e.g. Aizenman-Warzel \cite{AW2} or Kirsch \cite{K2}). However, on the Bethe tree Klein \cite{KleinB} proved
the existence of absolutely continuous spectrum  (see also Klein-Sadel \cite{KS}, Aizenman-Warzel \cite{AW}, Froese et. al. \cite{FHS}).

There are random Schr\"{o}dinger operators with decaying randomness for which a mobility edge is known to exist
(see Krishna \cite{MR1081712}, Kirsch et.al. \cite{KKO,MR1803388},  and Jaksic-Last \cite{JL4}.
These models are \emph{not} ergodic. However, the models we consider here are either ergodic in
$\ZZ^{\da}$-direction (for the strip) or even ergodic with respect to a $d$-dimensional sublattice.

\section{Setup}\label{sec:setup}
We consider quantum systems (wave guides) on $\xx\subset\ZZ^{d}=\ZZ^{d_{1}}\times\ZZ^{d_{2}}$.

For $p=(p_{1},\ldots,p_{d})\in(\NN\setminus\{1\})^{d}$ (periods) set (unit cell)
\begin{align}\label{eq:unitcell}
   \cc_{0}~=\{x\in\ZZ^{d}\mid 0\leq x_{\nu}\leq p_{\nu}-1 \text{ for all } \nu\}
\end{align}

By $e_{\nu}$ we denote the standard basis of $\ZZ^{d}$. The lattice $\LL$ and the subset $\LL_{M_{1}M_{2}}$
are defined by
\begin{align}
   \LL~&:=~\Big\{\sum_{\nu=1}^{d} i_{\nu}\,p_{\nu}e_{\nu}\mid i_{\nu}\in\ZZ\Big\}\\
\text{and}\qquad \LL_{M_{1}M_{2}}~&:=~\Big\{\sum_{\nu=1}^{d} i_{\nu}\,p_{\nu}e_{\nu}\mid M_{1}\leq i_{\nu}< M_{2}\text{ for } \nu\leq d_{1}\Big\}\,,
\end{align}
with $M_{1},M_{2}\in \ZZ,M_{1}< M_{2}$. In this paper we always assume that $d_{1},d_{2}>0$.

Then we define `cubes' $\xx_{M_{1}M_{2}}$ by `periodizing' $\cc_{0}$:
\begin{align}\label{eq:defxx1}
   \xx_{M_{1},M_{2}}~&:=~\cc_{0}+\LL_{M_{1}M_{2}}\\
   \intertext{and}
   \xx_{\infty}~&:=~\cc_{0}+\LL\,.\label{eq:defxx}
\end{align}
Informally, we consider $\xx_{\infty}$ as $\xx_{M_{1}M_{2}} $ with $M_{1}=-\infty, M_{2}=\infty$.

Thus $\xx_{M_{1}M_{2}} $ is a strip (`waveguide') of restricted width in $d_{1}$ directions and
unconfined in $d_{2} $ directions, $\xx_{\infty} $ is (for the moment) just a complicated expression for $\ZZ^{d} $.

Sometimes we omit the indices $M_{1},M_{2}$ and $\infty$ if they are clear from the context or if they are irrelevant.

The discrete Laplacian $H_{0}$ on $\xx_{\infty}=\ZZ^{d}$ is given by:
\begin{align}
   H_{0}\,u(n)~:=~\sum_{\nu=1}^{d} \Big(u(n+e_{\nu})\,+\,u(n-e_{\nu})\Big)
\end{align}
When we restrict $H_{0}$ to subsets of $\ZZ^{d}$ we have to impose boundary conditions.
In the following we will most of the time work with either `simple' boundary conditions
or with `periodic' boundary conditions.

\begin{defin}\label{def:sbc}
   If $\Lambda$ is a subset of $\ZZ^{d}$ then the operator $H_{0,\Lambda}$ on
$\ell^{2}(\Lambda) $ given by
\begin{align}
   H_{0,\Lambda}u(n)~=~\sum_{\nu=1}^{d} \Big(\chi_{\Lambda}(n+e_{\nu})\,u(n+e_{\nu})\;+\;\chi_{\Lambda}(n-e_{\nu})\,u(n-e_{\nu})\Big)
\end{align}
is called the Laplacian on $\Lambda $ with \emph{simple boundary conditions}.

Here
\begin{align}
\chi_{\Gamma}(n)=\left\{
                         \begin{array}{ll}
                           1, & \hbox{for $n\in\Gamma$;} \\
                           0, & \hbox{otherwise.}
                         \end{array}
                       \right.
\end{align}
\end{defin}

\begin{defin}\label{def:pbc}
Suppose the box $\Lambda\subset\ZZ^{d} $ is given by
\begin{align}
   \Lambda=\{ x\in\ZZ^{d}\mid q_{\nu}\leq x_{\nu}\leq p_{\nu} \text{ for }\nu=1,\ldots,d' \}
\end{align}
for some $d'\leq d$, then we call the operator $H_{0}^{\Lambda}$ defined by
\begin{align}
   H_{0}^{\Lambda}u(x)~=~\sum_{\nu=1}^{d}\, \big( u(N^{+}_{\nu}x)+u(N^{-}_{\nu}x)\big)
\end{align}
where
\begin{align}
   N^{+}_{\nu}x~&=~\left\{
                    \begin{array}{ll}
                      x+e_{\nu}, & \hbox{if $x+e_{\nu}\in\Lambda$;} \\
                      x-(p_{\nu}-q_{\nu}) e_{\nu}, & \hbox{if $x+e_{\nu}\notin\Lambda$.}
                    \end{array}
                  \right.\\
N^{-}_{\nu}x~&=~\left\{
                    \begin{array}{ll}
                      x-e_{\nu}, & \hbox{if $x-e_{\nu}\in\Lambda$;} \\
                      x+(p_{\nu}-q_{\nu}) e_{\nu}, & \hbox{if $x-e_{\nu}\notin\Lambda$.}
                    \end{array}
                  \right.
\end{align}
   the Laplacian on $\Lambda $ with periodic boundary conditions.
\end{defin}

For the operator $H_{0} $ on $\xx_{M_{1}M_{2}} $ we impose \emph{periodic} boundary conditions.

Now, we define the set $\Gamma $ of `active sites', i.~e. the sites where the potential $V_{\omega} $
may be nonzero.
The active sites \emph{inside} $\cc_{0} $ are denoted by $\Gamma_{0}$, with
\begin{align}\label{eq:assum1}
   \emptyset\neq \Gamma_{0}\varsubsetneq \cc_{0}
\end{align}
and
\begin{align}\label{eq:assum2}
   \Gamma~:=~~\Gamma_{0}+\LL_{M_{1}M_{2}}
\end{align}
where we include again the case $-M_{1}=M_{2}=\infty$.

\begin{exam}\label{exam:stand}
In the following examples $\xx$ may be either $\xx_{M_{1}M_{2}}$ or $\xx_{\infty}\;=\ZZ^{d}$
\begin{enumerate}
   \item For some $\nu\leq d_{1}$
\begin{align}\label{eq:singll}
   \Gamma=\{ x\in\xx\mid x_{\nu}=0 \}
\end{align}

   \item $\Gamma=\{ x\in\xx\mid x_{1}=0 \text{ or } x_{2}=0 \text{ or }\ldots
   \text{ or } x_{d_{1}}=0 \} $
   \item For some results we can deal with the following less restrictive model
   \begin{align}\label{eq:lrmodel}
      \emptyset~\not=~\Gamma~\subset~\{ x\in\xx\mid x_{1}=0 \text{ or } x_{2}=0 \text{ or }\ldots
   \text{ or } x_{d_{1}}=0 \}
   \end{align}
\end{enumerate}

\end{exam}

In this article we investigate spectral properties of operators $H$ on $\xx $ of the form
\begin{align}\label{eq:defh}
   H\,u(n)~=~H_{0}\,u(n)\;+\;V(n)\,u(n)
\end{align}
where the potential $V $ is supported by $\Gamma $, i.~e. $V(n)=0$ for $n\notin\Gamma $.

Most of the time we suppose that $V $ is a random potential with independent, identically distributed
random variables $V_{\omega}(\gamma),\gamma\in\Gamma$, but some of our results are independent of such an
assumption.

\section{Results}\label{sec:results}
We prove localisation under fairly weak assumptions on $\Gamma $.

\begin{thm}\label{thm:loc}
Suppose that $\emptyset\not=\Gamma\not=\xx $. Assume that the random variables $V_{\omega}(n), n\in\Gamma$ are independent with a common distribution $P_{0}$ which has a
bounded density $\rho $ (with respect to Lebesgue measure) with compact support.

If $I\subset\{ E\mid \dist(E,\sigma(H_{0,\Gamma^c})\geq \gamma \}$ then $H_{\omega}$ has pure point spectrum inside $I$ with exponentially decaying eigenfunctions if $\| \rho \|_{\infty} $ is small
enough, i.~e. if $\| \rho \|_{\infty} \leq c_{\gamma}$.
\end{thm}

\begin{cor}\label{cor:dloc}
   Under the assumptions of Theorem \ref{thm:loc} there is dynamical localisation in $I$.
\end{cor}

The following result shows that Theorem \ref{thm:loc} is not an empty statement.

\begin{prop}\label{prop:nonempty}
   Under the assumptions of Theorem \ref{thm:loc} there is an $\eta>0$ such that
\begin{align}
   [\inf \Sigma,\inf \Sigma+\eta] \cap \sigma(H_{0,\Gamma^{c}})~=~\emptyset\\
\text{and}\qquad [\sup \Sigma-\eta,\sup \Sigma] \cap \sigma(H_{0,\Gamma^{c}})~=~\emptyset
\end{align}
\end{prop}

Theorem \ref{thm:loc} and Corollary \ref{cor:dloc} are proved in Section \ref{sec:loc}. Proposition  \ref{prop:nonempty}. These results reprove and extend previous results
in \cite{JO,CR,EK,ES,KK3}.

Now, we turn to a class of examples for which we can prove the existence of \emph{absolutely
continuous spectrum}.

\begin{defin}\label{def:singlelayer}
  We call $\Gamma $ as in \eqref{eq:assum1} and \eqref{eq:assum2} a \emph{single layer set} if
$\Gamma\subset G\times \ZZ^{d_{2}}$ with
\begin{align}\label{cond:G0}
      \emptyset~\not=~G~\subset~G_{0}=\{ x\in\xx\mid x_{1}=0 \text{ or } x_{2}=0 \text{ or }\ldots
   \text{ or } x_{d_{1}}=0 \}
   \end{align}
A potential $W$ with $W(n)=0$ for $n\notin \Gamma$ for a single layer set $\Gamma $ is called a \emph{single
layer potential}.
\end{defin}
\begin{defin}\label{def:eL}
We set
\begin{align}
 \Ll~&:=~\bigtimes_{\nu=1}^{d_{1}}\;  \{0,1,2,\ldots,p_{\nu}-1\}\label{eq:Ll}\\
e_{L}~&:= 2\,\Big(\sum_{\nu=1}^{d_{1}}\,cos\big(\frac{\pi\ell_{\nu}}{p_{\nu}}y_{\nu}\big)\Big)\label{eq:eL}\\
\Ee~&:=~\bigcup_{L\in\Ll}~\;e_{L}+[-2d_{2},2d_{2}]\label{eq:Ee}
\end{align}
\end{defin}

\begin{thm}\label{thm:ac}
   Consider $H_{0}$ on $\xx_{M_{1}M_{2}}$ with $M_{1},M_{2}$ finite, $M_{2}-M_{1} $ even and with periodic boundary conditions. Assume that $\Gamma$ is a single layer set.

If $W$ is an arbitrary potential vanishing outside $\Gamma $, then

\begin{align}
   \Ee~\subset~\sigma_{ac}(H_{0}+W)\, .
\end{align}
\end{thm}

Theorem \ref{thm:ac} applies in particular to any $\Gamma$-trimmed random potential as in Theorem
\ref{thm:loc}.
Thus, for such a random potential there is an energy region with pure point spectrum
and a region with absolutely continuous spectrum. Consequently there exists a \emph{mobility edge}.

The following example is of particular interest:
\begin{prop}\label{pro:mobility}
   Suppose $d_{1}=1, d_{2}\geq 1 $ and take $\Gamma=p\ZZ\,\times\,\ZZ^{d_{2}} $ with $p\geq 2$.
Then
\begin{align}
   \sigma(H_{0,\Gamma^{c}})=\Ee\,.
\end{align}

\end{prop}
Hence, under the assumptions of Proposition \ref{pro:mobility} the $\inf \Ee $ and $\sup \Ee $ are
the mobility edges.

The proof of Theorem \ref{thm:ac} is contained in Section
\ref{sec:ac}.

\bigskip

Unfortunately, the proof of Theorem \ref{thm:ac} does not work for the case $-M_{1}=M_{2}=\infty$, i.~e. for $\xx=\ZZ^{d}$. However, for this case we can at least show, that both the fractional moment method and the multiscale analysis cannot work. In fact, the spectral values in $\Ee $ belong to
`extended states' in an informal sense.

Let $\Gamma\subset \xx_{\infty}$ be a one layer set and $V_{\omega}$ be a random potential on $\Gamma $
satisfying the assumptions of Theorem \ref{thm:loc}. Denote by $G_{E+i\zeta}^{V_{\omega}}(x,y)$
the Green function (i.~e. the kernel of the resolvent) with $\zeta>0 $.

Then we show
\begin{thm}\label{thm:green}We assume $\xx=\ZZ^{d}$ and $\Gamma$ is a single layer set.
\begin{enumerate}
   \item If $E\notin\sigma(H_{0,\Gamma^{c}})$ then for high enough disorder
\begin{align}
   \limsup_{\zeta\searrow 0}|G_{E+i\zeta}^{V_{\omega}}(x,y)|~\leq~ C\,e^{-m|x-y|}
\end{align}
$\PP$-almost surely.
\item If $E\in\Ee$ then
\begin{align}
   \limsup_{\zeta\searrow 0}\;\sum_{y\in\ZZ^{d}}|G_{E+i\zeta}^{V_{\omega}}(x,y)|~=~\infty
\end{align}
for all $x\notin\Gamma $ and all $\omega $.
\end{enumerate}

\end{thm}
Part 2 of Theorem \ref{thm:green} is actually a deterministic result, it holds for \emph{any}
potential vanishing outside $\Gamma $.

We prove this theorem in Section \ref{sec:absence}.

\section{The Random Operator and its Spectrum}\label{sec:random}
In this section we consider operators with \emph{random} potential. To emphasize this we
write
\begin{align}\label{eq:homega}
   H_{\omega}~:=~H_{0}\,+\,V_{\omega}\, .
\end{align}

\begin{hyp}\label{hyp:potential1}
We suppose that the potentials
$V_{\omega}(\gamma),\gamma\in\Gamma$ are i.i.d. with a common distribution $P_{0}$.
We assume that the support $\WW$ of $P_{0} $ is compact.
\end{hyp}
The assumption that $\WW $ is compact can be relaxed considerably but we don't bother to do so.

We denote the corresponding probability space by $(\Omega,\ff,\PP) $ which can and will be taken to be
$\Big(\WW^{\Gamma}, \bigotimes_{\gamma\in\Gamma}\bb(\WW),\bigotimes_{\gamma\in\Gamma} P_{0}\Big) $.

We write the lattice $\LL $ as $\LL=\LL_{1}\times\LL_{2} $ with $\LL_{1}\subset\ZZ^{d_{1}}$ and
$\LL_{2}\subset\ZZ^{d_{2}}$ and points $x$ in $\xx$ as $x=(x_{1},x_{2})$ with $x_{i}\in\ZZ^{d_{i}}$.

We define `shift' operators $T_{j}, j\in \LL':=\LL_{2}$ on $(\Omega,\ff,\PP) $ by
\begin{align}
   T_{j}\omega(x_{1},x_{2})~:=~\omega(x_{1},x_{2}-j)
\end{align}

It is easy to see that the shift $T_{j} $ is measure preserving, i.~e. $\PP({T_{j}}^{-1}A)=\PP(A) $ for every $A\in\ff $.

The following result tells us that the family $\{T_{j}\}_{j\in\LL'}$ is \emph{ergodic}:
\begin{prop}\label{prop:ergodic}
   If $A\in\ff$ is invariant under $\{T_{j}\}_{j\in\LL'} $, i.~e. ${T_{j}}^{-1}A=A $ for all
   $j\in\LL'$, then either $\PP(A)=0 $ or
   $\PP(A)=1 $.
\end{prop}
This result can be found in  \cite{EKSS} for example.

Define for $j\in\LL'$ the shift operator
\begin{align}
U_{j}u(x_{1},x_{2}):=u(x_{1},x_{2}-j)\,,
\end{align}
for $(x_{1},x_{2})\in\xx$.

The operators $U_{j}$ are unitary on $\ell^{2}(\xx)$ , moreover the operators $H_{\omega}$ are \emph{ergodic} in the sense
\begin{align}
   H_{T_{j}\omega}~=~U_{j}H_{\omega}U_{j}^*\,.
\end{align}
with ergodic $T_{j} $ by Proposition \ref{prop:ergodic}

It follows (see e.~g. \cite{KMCrelle}):
\begin{prop}\ \\[-5mm]
\begin{enumerate}
   \item The spectrum $\sigma(H_{\omega})$ is non random (almost surely).
   \item The same is true for the measure theoretic parts of the spectrum (the absolutely continuous part
   $\sigma_{ac}(H_{\omega})$, the singular continuous part etc.).
   \item There is (almost surely) no discrete spectrum.
\end{enumerate}
\end{prop}

\begin{defin}
   We denote by $\Sigma $ the almost sure spectrum of $H_{\omega} $, i.~e.\goodbreak $\Sigma=\sigma(H_{\omega})$
   $\PP $-almost surely.
\end{defin}

We now investigate the spectrum (as a set).
\begin{defin}
   A function $W:\xx\in\RR$ is called an \emph{admissible potential} (with respect to $P_{0}$) if
   \begin{align*}
      W(x)~&\in \WW=\supp P_{0} &&\rm{if }\;x\in \Gamma,\\
      W(x)~&=~0 &&\rm{otherwise.}
   \end{align*}
    We denote the set of admissible potentials by $\aa$.
\begin{rem}\label{rem:admissible}
   Taking
   \begin{align}
      \big(\Omega,\ff,\PP\big)~=~ \Big(({\WW})^{\Gamma}, \bigotimes_{\gamma\in\Gamma}\bb(\WW ),\bigotimes_{\gamma\in\Gamma} P_{0}\Big)
   \end{align}
  there ia a one-to-one correspondence $\tau $ between $\Omega $ and the set $\aa $ of admissible
  potentials, namely $\tau(\omega)(n)=\sum_{\gamma\in\Gamma} \omega_{\gamma} \delta_{\gamma\, n} $.

  We may therefore identify $\Omega $ and $\aa $.
\end{rem}
\end{defin}
\begin{thm}
\begin{enumerate}
   \item If $W$ is an admissible potential then $\sigma(H_{0}+W)\subset\Sigma$.
    \item We have \begin{align}
             \Sigma~=~\bigcup_{W\in\aa}\sigma(H_{0}+W)\,.
          \end{align}
\end{enumerate}
\end{thm}
\begin{proofn}
1. For $E\in\sigma(H_{0}+W) $ there exists a Weyl sequence of functions $\varphi_{n} $ with
compact (hence finite) support, more precisely we may suppose that $\| \varphi_{n} \|=1 $ and
\begin{align}
   \| (H_{0}+W\,-E)\,\varphi_{n}\|~<~\frac{1}{n}
\end{align}
Set $S_{n}=\supp \varphi_{n}$ which is a finite set. By the Borel-Cantelli Lemma there is a vector $j_{n}\in\LL'$
such that
\begin{align}
   \sup_{k\in S_{n}}\;|\,W(k)-V_{\omega}(k+j_{n})\,|~<~\frac{1}{n}
\end{align}
With $\psi_{n}(x)=\varphi_{n}(x+j_{n}) $ we therefore get
\begin{align}
   \|\, (H_{\omega}\,-E)\,\psi_{n}\,\|~<~\frac{2}{n}\,,
   \end{align}
thus $E\in\sigma(H_{\omega})$.

2. Since the set $\aa $ has probability one (in the sense of Remark \ref{rem:admissible})
\begin{align}
             \Sigma~\subset~\bigcup_{W\in\aa}\sigma(H_{0}+W)\,.
          \end{align}
This together with 1. proves the theorem.
\end{proofn}

Set $V_{a}=a \chi_{\Gamma}$ then $V_{a}$ is an admissible potential if $a\in\supp P_{0}$.
We define $H_{a}=H_{0}+V_{a}$
and set $E_{min}(a)=\inf\sigma(H_{a})$ and $E_{max}(a)=\sup\sigma(H_{a})$.

\begin{thm}
   If $\supp P_{0}=[a,b]$ then
\begin{align}
   \Sigma~=~[E_{min}(a),E_{max}(b)]\,.
\end{align}
\end{thm}
\begin{proofn}
  Since $V_{x}$ are admissible for all $x\in[a,b]$ and due to continuity we have  $\Sigma\supset[E_{min}(a),E_{max}(b)] $.

Suppose now, that $W$ is an admissible potential then by monotonicity $$H_{a}\leq H_{0}+W\leq H_{b}\,.$$
So, $\sigma(H_{0}+W)\subset [\inf\sigma(H_{a}),\sup\sigma(H_{b})]$.
\end{proofn}

We will have a closer look at the operators $H_{a}$. Let us denote by $H_{0,\Gamma}$ and
$H_{0,\Gamma^{c}} $ the operator $H_{0}$ restricted to $\ell^{2}(\Gamma) $ and
$\ell^{2}(\Gamma^{c}) $ respectively with simple boundary conditions.

\begin{prop} \label{prop:positive} For any $a\in\RR$
   \begin{align}
     \inf \sigma(H_{a})~&<~\inf \sigma(H_{0,\Gamma^{c}})~\label{eq:Ha1}\\
\text{and}\qquad~\sup \sigma(H_{a})
~&>~\sup \sigma(H_{0,\Gamma^{c}})\label{eq:Ha2}
   \end{align}
\end{prop}
\begin{proofn}
   Take $\varphi\in\ell^{2}(\Gamma^{c}) $ with $\| \varphi \|_{\ell^{2}(\Gamma^{c})} =1$ and define
$\widetilde{\varphi}\in\ell^{2}(\xx) $ by
   \begin{align}
      \widetilde{\varphi}(n)~=~\left\{
                                 \begin{array}{rl}
                                   \varphi(n), & \hbox{for $n\in\Gamma^{c} $;} \\
                                   0, & \hbox{otherwise.}
                                 \end{array}
                               \right.
   \end{align}
Then
\begin{align}\label{eq:sp}
   \langle \varphi,H_{0,\Gamma}\, \varphi\rangle_{\ell^{2}(\Gamma^{c})}~
=~\langle \widetilde{\varphi},H_{a}\widetilde{\varphi}) \rangle_{\ell^{2}(\xx)}\,.
\end{align}
From \eqref{eq:sp} the equalities \eqref{eq:Ha1} and \eqref{eq:Ha2} follow by the Min-Max-Principle in their
\emph{non strict} version ($\leq$ and $\geq$).

To prove the strict inequalities we show that $\inf H_{a}$ is strictly increasing with $a$.

The operator $H_{a}$ is periodic, therefor its ground state energy $E_{a}$ is given by the ground state $\psi_{a} $ of the operator $h_{a}$ on the periodic cell $C_{0}$ with periodic boundary conditions.
The eigenfunction $\psi_{a} $ is strictly positive and $h_{a}\psi_{a}=E_{a}\psi_{a}$.

By the Hellmann-Feynman Theorem
\begin{align}
   \frac{d}{da}E_{a} ~=~\sum_{j\in C_{0}\cap \Gamma}\,|\psi_{a}(j)|^{2}~>~0\,,
\end{align}
hence $E_{a}<E_{b}$ if $a<b$.
This proves the strict version of \ref{eq:Ha1}. \eqref{eq:Ha2} is proved similarly.

\end{proofn}

Now, we consider the case that $\Gamma=G\times \ZZ^{d_{2}}$. In this case the operator $H_{a}$
separates in the following way.
\begin{defin}\label{def:split}
  The space $\xx $ splits in a part $\zz\subset\ZZ^{d_{1}}$ and $\ZZ^{d_{2}} $, namely
\begin{align}
   \xx_{\infty}~&=~\ZZ^{d_{1}}\times\ZZ^{d_{2}}\\
\xx_{M_{1}M_{2}}~&=~\zz\times\ZZ^{d_{2}}
\end{align}

We denote  the Laplacian on  $\ell^{2}(\zz)$ (possibly with periodic boundary conditions) by $H^{(1)}_{0}$ and the Laplacian on  $\ell^{2}(\ZZ^{d_{2}})$ by $H^{(2)}_{0}$.
We also set $H^{(1)}_{a}=H^{(1)}_{0}+a\chi_{G}$.

\end{defin}

Then
\begin{align}
   H_{a}~=~\big(H_{a}^{(1)}\otimes \mathbf{1}_{\ZZ^{d^{2}}}\big)\;\oplus\;\big(\mathbf{1}_{\zz}\otimes H_{0}^{(2)}\big)
\end{align}

 Consequently
$\sigma(H_{a})=\sigma(H^{(1)}_{a})+[-2d_{2},2d_{2}]$.

This proves the following Corollary:
\begin{cor}
If $\Gamma=G\times \ZZ^{d_{2}} $ and $\supp P_{0}=[a,b]$ then
  \begin{align}
   \Sigma~=~[\inf\sigma(H^{(1)}_{a}),\sup\sigma(H^{(1)}_{b})]\;+\;[-2d_{2},2d_{2}]
\end{align}
\end{cor}

\section{Localisation}\label{sec:loc}
In this section we prove pure point spectrum with exponentially decaying eigenfunction for energies
$E\notin\sigma(H_{0,\Gamma^{C}})$ for high enough disorder.

For the ($\LL$-periodic) set $\Gamma $ of `active' sites we merely assume that $\Gamma\not=\emptyset $.
We may also assume that $\Gamma\not=\cc_{0} $, since otherwise $\Gamma $ is the whole space, a case
which is known, of course.

In the following we consider the case $\xx=\xx_{\infty} $, at the end of this section we comment
on the case $\xx_{M_{1}M_{2}} $.

We will use multiscale analysis (see \cite{K2} and references given there).
During the proof we need to consider boxes $\Lambda $ which are unions of shifted $\cc_{0}$.
Recall that the unit cell $\cc_{0}$ is defined by:
\begin{align}
   \cc_{0}~=~\{x\in\ZZ^{d}\mid 0\leq x_{\nu}\leq p_{\nu}-1 \text{ for all } \nu\}
\end{align}

\begin{defin}
   We call a set $\Lambda\subset\ZZ^{d} $ a $\cc_{0}$-box, if
\begin{align}
   \Lambda~=~\{x\in\ZZ^{d}\mid L_{\nu}p_{\nu}\leq x_{\nu}\leq L_{\nu}' p_{\nu}-1 \text{ for all } \nu\}
\end{align}
\end{defin}
One of the crucial
ingredients of (most versions of) multiscale analysis is the Wegner-Estimate. To prove this we need the following result.
By $H_{\omega}^{\Lambda} $ we denote the operator $H_{\omega}=H_{0}+V_{\omega}$ restricted
to $\ell^{2}(\Lambda)$ with periodic boundary conditions.
\begin{prop}\label{prop:uc}
Suppose $\Lambda $ is a $\cc_{0}$-cube and $E\notin\sigma(H_{0,\Gamma^{c}})$ and
\begin{align}\label{eq:ev}
   H_{\omega}^{\Lambda}\psi~=~E\psi
\end{align}
then
\begin{align}
   \| \psi \|_{\ell^{2}(\Lambda)}~\leq~\frac{C}{\dist \big(E,\sigma(H_{0,\Gamma^{c}})\big)}\;
\| \psi \|_{\ell^{2}(\Lambda\cap \Gamma)}
\end{align}
with a constant $C $ which is independent of $\Lambda$ and $E$.

\end{prop}
\begin{proofn}
Set $\Lambda_{1}=\Lambda\cap \Gamma^{c} $ and $\Lambda_{2}=\Lambda\cap \Gamma $.
   We write $\ell^{2}(\Lambda) $
\begin{align}
   \ell^{2}(\Lambda)~=~\ell^{2}(\Lambda_{1})\,\oplus\,\ell^{2}(\Lambda_{2})
\end{align}
Accordingly, we may write the operator $H_{\omega}$ in block matrix form
\begin{align}
   H_{\omega}^{\Lambda}~=~\begin{pmatrix} H_{0}^{\Lambda_{1}} & T\\ T^{*} & H_{0}^{ \Lambda_{2}}+V_{\omega} \end{pmatrix}
\end{align}
The operator $T:\ell^{2}(\Lambda_{2})\to\ell^{2}(\Lambda_{1})$ `restores' the links between $\Lambda_{2} $ and $\Lambda_{1} $.

The eigenvalue equation \eqref{eq:ev} reads
\begin{align}\begin{pmatrix} H_{0}^{\Lambda_{1}} & T\\ T^{*} & H_{0}^{ \Lambda_{2}}+V_{\omega} \end{pmatrix}
   \begin{pmatrix}
      \psi_{1}\\\psi_{2}
   \end{pmatrix}
~=~E\;\begin{pmatrix}
      \psi_{1}\\\psi_{2}
   \end{pmatrix}
\end{align}
So, in particular
\begin{align}
   (H_{0 }^{\Lambda_{1}}-E)\,\psi_{1}~=~-T\,\psi_{2}
\end{align}
Thus
\begin{align}
   \| \psi \|_{\ell^{2}(\Lambda_{1})}~&=~\| \psi_{1} \|\\
&\leq~\| (H_{0}^{\Lambda_{1}}-E)^{-1}\|\,\| T \|\,\| \psi_{2} \|\label{eq:est}
\end{align}
Since for given $m$ we have  $|T(n,m)|=1$ for at most $2d$ point $n$ and
$T(n,m)=0$ otherwise, we conclude  $\| T \|\leq 2d $.

Since we consider periodic boundary conditions on $\Lambda $ we have
\begin{align}
   \sigma(H_{0}^{\Lambda_{1}})~\subset~\sigma(H_{0,\Gamma^{c}})
\end{align}
as any eigenfunction on $\Lambda $ can be periodically extended to an eigenfunction on $\xx $.
\end{proofn}

Now we turn to the Wegner estimate. By $N(A,E)$ we denote the number of
eigenvalues of the operator $A$ up to $E$.

\begin{thm} If $\dist\big(E,\sigma(H_{0,\Gamma^{c}})\big)\geq \gamma$ and $0\leq \varepsilon\leq \half\gamma $
then
   \begin{align}\label{eq:Wegner}
      \EE\Big(N\big(H_{\omega}(\Lambda),E+\varepsilon\big)-N\big(H_{\omega}(\Lambda),E-\varepsilon\big)
\Big)~\leq~\frac{C}{\gamma}\,\| \rho \|_{\infty}\;\varepsilon\;|\Lambda|
   \end{align}
where $|\Lambda|$ denote the volume (number of points) of $\Lambda $.
\end{thm}

\begin{proofn}
The proof is a combination of the proofs from \cite{K2} and \cite{K94}.
We sketch the main ideas. We use the abbreviation $H=H_{\omega}(\Lambda)$.

Let $g $ be a monotone $C^{\infty} $-function with $0\leq g(t) \leq 1$, $g(t)=0 $ for $t\leq 2\varepsilon $
and $g(t)=1 $ for $t\geq 2\varepsilon $.

We obtain
\begin{align}
   N(H,E+\varepsilon)-N(H,E+\varepsilon)~&\leq~\tr g(H-E+2\varepsilon)\,-\,\tr g(H-E-2\varepsilon)\\
   &=~\int_{E-2\varepsilon}^{E+2\varepsilon} \tr g'(H-\lambda)\,d\lambda
\end{align}

Let $E_{n}$ denote the eigenvalues of $H_{\omega}(\Lambda)$ labelled in increasing order.
These eigenvalues depend on the values $v_{j}:=V_{\omega}(j), j\in\Lambda_{2}$.

Thus we may consider
\begin{align}
   \sum_{j\in\Lambda_{2}} \frac{\partial}{\partial v_{j}} \tr g(H-\lambda)~&=~\sum_{n}\sum_{j\in\Lambda_{2}} \frac{\partial}{\partial v_{j}} g(E_{n}-\lambda)\\
&=~\sum_{n} g'(E_{n}-\lambda)
\sum_{j\in\Lambda_{2}}\frac{\partial E_{n}}{\partial v_{j}}\\
&\geq~\sum_{n} g'(E_{n}-\lambda)\;\sum_{j\in\Lambda_{2}}\,|\psi_{n}(j)|^{2}\\
&\geq~{C'}\,{\dist\big(E,\sigma(H_{0, \Gamma^{c}})\big)}\;\tr g'(H-\lambda)
\end{align}
where $\psi_{n} $ is a normalised eigenfunction of $H$ with eigenvalue $E_{n}$. Above we used the Hellmann-Feynman Theorem and, in the final step, Proposition \ref{prop:uc}.

Summarising we proved
\begin{align}
  &\EE\Big(N(H,E+\varepsilon)-N(H,E+\varepsilon)\Big)\\~\leq~
&\frac{C''}{\dist\big(E,\sigma(H_{0 \Gamma^{c}})\big)}\sum_{j\in\Lambda_{2}} \int_{E-2\varepsilon}^{E+2\varepsilon} \EE\Big(\frac{\partial}{\partial v_{j}} \tr g(H-\lambda)\Big)\,d\lambda
\end{align}

Suppose $\supp P_{0}\subset [a,b]$ and denote by $H(v_{j}=c)$ the operator $H$ with $V_{j}$ replaced by the value $c$, then
\begin{align}
   &\int \frac{\partial}{\partial v_{j}} \tr g(H-\lambda)\rho(v_{j})\,dv_{j}\\
~\leq~
&\| \rho \|_{\infty}\,\Big(\tr g\big(H(v_{j}=b)-\lambda\big)-\tr g\big(H(V_{j}=a)-\lambda\big)\Big)
\leq \|~ \rho \|_{\infty}
\end{align}
We used that changing the potential at one site $j$ is a rank one perturbation and $0\leq g(\lambda)\leq 1$.

Performing the integrals over the $v_{k}, k\not=j$ gives the desired result.

\end{proofn}

Once we have the Wegner estimate the multiscale analysis follows the usual path. We need
an initial length scale estimate and the induction step over growing length scales.

The initial length scale estimate follows directly from the Wegner estimate \ref{eq:Wegner}. As long as we are away from the spectrum of $H_{0,\Gamma^{c}}$ we can make the right hand side as small as we
like by taking $\| \rho \|_{\infty} $ small. This corresponds to high disorder.

The induction step follows the lines in \cite{K2} sections 9 and 10. The only difference being
that we deal with periodic boundary conditions while \cite{K2} uses simple boundary conditions.

If $\xx=\xx_{M_{1}M_{2}} $ we start the induction with a cube of the form
\begin{align}
   \Lambda=\{ x\in\xx_{M_{1}M_{2}}\mid L_{\nu}p_{\nu}\leq x_{\nu} \leq L_{\nu}'p_{\nu}-1 \text{ for } \nu=d_{1}+1,\ldots,d_{1}+d_{2} \}
\end{align}

Corollary \ref{cor:dloc} follows from the work \cite{DamanikS} of Damanik and Stollmann.

\section{Absolutely Continuous Spectrum}\label{sec:ac}

In this section we consider a special case of the above operators.

We start with the following observation:

\begin{prop}\label{prop:specW} Assume $\Gamma=G\times \ZZ^{d_{2}}$ and
suppose the (otherwise arbitrary) potential $W$ is concentrated on $\Gamma $. If
    there exists a polynomially bounded solution $\psi$ of
\begin{align}
   H^{(1)}_{0}\psi~=~e\,\psi
\end{align}
which vanishes on $G $,
then
\begin{align}
   e\,+\,[-2d_{2},2d_{2}]~\subset~\sigma(H_{0}+W)
\end{align}
\end{prop}
\begin{rem}
   $H_{0}^{(1)}$ and $H_{0}^{(2)}$ were defined in Definition \ref{def:split}.
\end{rem}
\begin{proofn}
Any $\eta\in[-2d_{2},2d_{2}]$ is of the form $\eta=2\sum_{\nu=1}^{d_{2}}\cos(\kappa_{\nu})$ and
$\varphi(x)=\prod_{\nu=1}^{d_{2}} \sin(\kappa_{\nu}x_{\nu}) $ is a (bounded) function solving
\begin{align}
   H^{(2)}_{0}\varphi~=~\eta\,\varphi\,.
\end{align}
This can be verified by applying the addition theorem for the sinus.

Consequently, $\Psi(x,y):=\psi(x)\,\varphi(y)$ is a bounded solution to
\begin{align}
   H_{0}\,\Psi~=~(e+\eta)\,\Psi\,.
\end{align}
Since $\psi $ vanishes on $G $, $\Psi $ vanishes on $\Gamma $, so
\begin{align}
   (H_{0}+W)\,\Psi~=~H_{0}\,\Psi~=~(e+\eta)\,\Psi\,.
\end{align}
Thus, $e+\eta$ is a generalized eigenvalue of $H_{0}+W$.
By Sch'nol's Theorem any generalized eigenvalue belongs to the spectrum (see \cite{MR670130}, Section C4 or \cite{Invitation}, Section 7.1).
\end{proofn}

We discuss a class of examples for which Proposition \ref{prop:specW} applies.

We look at $\xx $ or at the strip $\xx_{M_{1}M_{2}} $. In the latter case we impose periodic boundary conditions and take $M_{2}-M_{1} $ is even. This way we avoid to discuss various cases separately.

For $L=(\ell_{1},\ldots,\ell_{d_{1}}), \ell_{\nu}\in \{1,2,\ldots,p_{\nu}-1\}$ we set
\begin{align}
   \Phi_{L}(x_{1},\ldots,x_{\da})~:=~\prod_{\nu=1}^{\da}\;\sin\Big(\frac{\pi\ell_{\nu}}{p_{\nu}}x_{\nu}\Big)
\end{align}
\begin{lemma}
   Under condition \eqref{cond:G0} the functions $\Phi_{L}(x) $
      is a solution to
   \begin{align}\label{eq:solution}
      (H^{(1)}_{0}+W)\psi~=~2\,\Big(\sum_{\nu=1}^{\db}\,cos\big(\frac{\pi\ell_{\nu}}{p_{\nu}}x_{\nu}\big)\Big)\;\psi
   \end{align}
and
\begin{align}
   \Phi_{L}(x)~=~0 \qquad\qquad \text{for }\;x\in G
\end{align}
\end{lemma}
\begin{proofn}
   Again by applying addition theorems and the fact that $\Phi_{L} $ vanishes on $G_0$, hence
   on $G $, we see that
   \eqref{eq:solution} holds. Moreover, since $M_{2}-M_{1} $ is even $\Phi_{L} $ satisfies periodic
   boundary conditions.
\end{proofn}
\medskip

Now, we are ready to prove Theorem \ref{thm:ac}.
\bigskip

\begin{proofn}[ (Theorem \ref{thm:ac})\,]
   Take $E\in\Ee $ then $E=e_{L}+\eta $ for some $\eta\in[-2\db,2\db] $. Denote by $E_{L}$ the
   (one dimensional) subspace of $\ell^{2}(\zz) $ generated by the eigenfunction $\Phi_{L}$.

   The (closed) subspace $\mathfrak{h}_{L}=E_{L}\otimes\ell^{2}(\ZZ^{\db}) $ of $\ell^{2}(\xx) $ is invariant under the operator $H_{0}+W$
   and restricted to $\mathfrak{h}_{L}$ the operators $H_{0}+W$ and $H_{0} $ agree. Consequently,
   $H_{0}+W$ on $\mathfrak{h}_{L}$ is unitarily equivalent to $H_{0}^{(2)}+e_{L}$ on
   $\ell^{2}(\ZZ^{\db}) $, an operator with purely absolutely continuous spectrum.

\end{proofn}

\section{Absence of Exponential Localisation}\label{sec:absence}

We start with a general observation.
Let $W$ be an arbitrary potential and denote by $G^{W}_{z}(x,y) $ the Green's function for $H:=H_{0}+W$, i.~e.  the kernel of the operator $(H_{0}+W-z)^{-1}$.

\begin{thm}\label{thm:noexp}
If for some $E\in\RR $
\begin{align}
   H\,\psi~=~\big(H_{0}+W\big)\,\psi~=~E\,\psi
\end{align}
for a bounded function $\psi $, then for all $x\in\ZZ^{d} $ with $\psi(x)\not=0 $

\begin{align}
   \liminf_{\zeta\searrow 0}\;\zeta\,\sum_{y\in\ZZ^{d}}|G^{W}_{E+i\zeta}(x,y)|~>~0\,,
\end{align}
in particular
\begin{align}\label{eq:noexp}
   \sup_{\zeta\searrow 0}\sum_{y\in\ZZ^{d}}|G^{W}_{E+i\zeta}(x,y)|~=~\infty\,.
\end{align}
\end{thm}

Observe that $|G_{E+i\zeta}(x,y)|\leq \frac{C}{\eta}\,e^{-c\eta\| x-y \|} $ by
the Combes-Thomas estimate (see e.~g. \cite{K2}). Thus for any $\zeta>0 $
\begin{align}
   \sum_{y\in\ZZ^{d}}|G^{W}_{E+i\zeta}(x,y)|~<~\infty\,.
\end{align}

\begin{proofn}[]
Take $\varepsilon>0 $ arbitrary and assume that $|\psi(x)|\leq A <\infty $.

Let
\begin{align*}
   \Lambda_{L}~&=~\{n\in\ZZ^{d}\mid |n_{\nu}|\leq L \text{ for }\nu=1,\ldots,d \}\\
\text{and }\quad\partial\,' \Lambda_{L}~&=~\{ n\in\ZZ^{d}\mid |n_{\nu}-L|\leq 2 \text{ for some }\nu \}
\end{align*}

Denote by $\chi_{L} $ the characteristic function of $\Lambda_{L} $ and
set $\psi_{L}:=\chi_{L}\,\psi$.

We compute
\begin{align}
   H\,\psi_{L}(x)~&=~\chi_{L}(x)\big(H\psi\big)(x)\;
+\,\sum_{|j|=1}\psi(x+j)\big(\chi_{L}(x+j)-\chi_{L}(x)\big)\notag\\
&=~E\,\psi_{L}(x)\;+\;\sum_{|j|=1}\psi(x+j)\big(\chi_{L}(x+j)-\chi_{L}(x)\big)\notag\,.
\end{align}
It follows that for $L$ big enough
\begin{align}
   \psi(x)=\sum_{y\in \ZZ^{d}}G^{W}_{E+i\zeta}(x,y)~
\Big(\sum_{|e|=1}\psi(y+e)\big(\chi_{L}(y+e)-\chi_{L}(y)\big)\,-i\zeta \psi(y)\chi_{L}(y)\Big)\notag
\end{align}
Observe that $ \chi_{L}(x+j)-\chi_{L}(x)$ (with $|j|=1 $) vanishes outside $\partial\,'\Lambda_{L} $.
Consequently
\begin{align}
   |\psi(x)|~\leq~2d\,A\,\sum_{y\in \partial\,'\Lambda_{L}}\,|G^{W}_{E+i\zeta}(x,y)|~+
~\zeta \,\sum_{y\in \ZZ^{d}}\,|G^{W}_{E+i\zeta}(x,y)|
\end{align}
Since $\sum_{y\in \ZZ^{d}}\,|G^{W}_{E+i\zeta}(x,y)|<\infty $ for $\zeta>0 $ we can
choose $L$ (depending on $\zeta>0 $) such that
$2d\,A\,\sum_{y\in \partial\,'\Lambda_{L}}\,|G^{W}_{E+i\zeta}(x,y)|<\varepsilon $
Then
\begin{align}\label{eq:psiest}
   |\psi(x)|~\leq~\zeta \,\sum_{y\in \ZZ^{d}}\,|G^{W}_{E+i\zeta}(x,y)|~+~\varepsilon
\end{align}
Now suppose that $\liminf_{\zeta\nearrow 0} \zeta \,\sum_{y\in \ZZ^{d}}\,|G^{W}_{E+i\zeta}(x,y)|=0$
Then as $\varepsilon $ was arbitrary \eqref{eq:psiest} implies $\psi(x)=0 $ which is a contradiction.

\end{proofn}

We apply the above theorem to our model.
\begin{thm}\label{thm:noexpmodel}
  Assume Condition \eqref{cond:G0}
    holds and let $W $ be an admissible potential on $\xx_{\infty} $.

Then for each $x_{0}\in\ZZ^{d}$ and for all $E\in\Ee $
\begin{align}\label{eq:noexp1}
  \sup_{x\in x_{0}+\cc_{0}} \sup_{\zeta\searrow 0}\sum_{y\in\ZZ^{d}}|G^{W}_{E+i\zeta}(x,y)|~=~\infty
\end{align}
\end{thm}
\begin{proofn}[]

Take $E\in\Ee$, then $E=e_{L}+\eta$ with
\begin{align}
   \eta~=~2\sum_{k=1}^{\db} \cos(\pi \kappa_{k})
\end{align}
for some $L $ and some $\kappa_{k} $.
It follows that
\begin{align}
   \psi(x_{1},\ldots,x_{\da},y_{1},\ldots,y_{\db})~:=~\prod_{j=1}^{\da}\,\sin(\frac{\pi \ell_{j}}{p_{j}} x_{j} )
    ~\prod_{k=1}^{\db}\, e^{\rm{i}\pi \kappa_{k} y_{k}}
\end{align}
vanishes on $\Gamma $ and is a solution to
\begin{align}
   H_{0}\psi~=~\big(H_{0}+W\big) \psi~=~E\;\psi\,,
\end{align}
with $||\psi||_{\infty}\leq 1 $.

An application of Theorem \ref{thm:noexp} gives the result.
\end{proofn}

\bibliographystyle{plain}
\bibliography{Waveguides}
\end{document}